\documentclass[9pt,pra,noshowpacs,preprintnumbers,amsmath,amssymb,reprint,twocolumn]{revtex4}
\usepackage[utf8]{inputenc}
\usepackage{graphicx}% Include figure files
\usepackage[colorlinks=true,linkcolor=blue,citecolor=blue,urlcolor=red]{hyperref}

\usepackage{siunitx}
\DeclareSIUnit\bar{bar}
\DeclareSIUnit\angstrom{\text {Å}}

\begin{document}
\title{Index-leveling for forced-flow turbulent face-cooling of laser amplifiers}

\author{Denis Marion\textsuperscript{1}, Philippe Balcou\textsuperscript{1}, Christophe Féral\textsuperscript{1}, Antoine Rohm\textsuperscript{1} and Jérôme Lhermite\textsuperscript{1}}

\affiliation{\textsuperscript{1}Universit\'e de Bordeaux, CNRS, CEA, CELIA (Centre Lasers Intenses et Applications), UMR 5107, 33405 Talence, France}

%\email{denis.marion@u-bordeaux.fr} 

\begin{abstract}
Direct laser slab face-cooling by a fluid crossing the main and pump laser beams is an important method to reach high average laser powers. However, the flow regime is usually maintained at low Reynolds numbers, to prevent the onset of turbulence features in the flow, that would degrade wavefront quality. We show here how bringing the fluid temperature to the thermo-optical null point, close to the water/ice transition in the case of water, allows one to mitigate the optical consequences of hydrodynamic instabilities, by bleaching optically the temperature inhomogeneities within the flow. 
This optical process, dubbed \emph{index-leveling}, opens the door to a highly efficient forced-flow, weakly turbulent face-cooling regime that should be instrumental to boost the kilowatt capabilities of next generation high-power lasers.
\end{abstract}

\framebox[.975\textwidth][c]{
\href{https://doi.org/10.1364/OL.455616}{DOI: 10.1364/OL.455616}
\hspace*{\fill}Please cite as: Opt. Lett. \textbf{47}, 2850-2853 (2022)
}

\maketitle

Average powers of research-class lasers, such as used in ultra-high intensity laser matter interactions, have increased steadily in the last few years, thanks to few key technologies, such as Large Mode Fibers \cite{Richardson:10}, thin disk technologies \cite{Giesen:07}, and cryogenic helium cooling \cite{Albach:19}, leading however to severe problems of thermal management.
 Indeed, a large fraction of the power deposited by pump lasers is downgraded into heat \cite{Lin:17}, which has to be efficiently removed, otherwise the laser medium heats up, thus reducing amplification, and inducing potentially catastrophic internal stress leading to wavefront deformation, birefringence, and ultimately medium breakdown. 
 
 An efficient cooling method makes use of a cooling liquid that flows directly onto the surface of the amplifier slab to extract the  heat \cite{Okada:06,lhermite:21}. In this scheme the coolant is crossed by  the pump laser beams and by the amplified laser, leading to concerns for the wavefront quality after propagation through a liquid that is subject to complex hydrodynamic and thermal processes.
 
 Different cooling liquids have been proposed, such as heavy water\cite{Siebold:14}, organic solvents and special oils; however de-ionized water remains the most common coolant, thanks to the ease of use, low price and large thermal capacity of water.
 Many studies have also been devoted to the hydrodynamics of water under thermal charges, or optical transmission in turbulent media\cite{Hill:78,Hill:81}, allowing the laser physicist to benefit from decades-long insights by physicists of fluid mechanics.
 In the usual case of no external heat input, the flow is known to occur in either the laminar regime, with smooth thermo- and hydrodynamic fields, or in the turbulent regime. Transition between laminar and turbulent regimes occurs when the dimensionless Reynolds number Re, whose significance will be recalled in the following, reaches values of the order of 2300.
 The regularity and homogeneity of low-Re flows allow one to cool with negligible wavefront distortions, but those exhibit low heat exchange coefficients with the slab.
 Flows with higher Re are in a transition, or weak turbulence regime, up to $\text{Re}=10000$ where strong turbulence sets in.
 High-Re flows display higher exchange coefficients, but when a turbulent fluid is in contact with a heating surface,  unstable fluid elements with different temperatures appear randomly, leading to inhomogeneities in temperatures and hence in optical indices $n$, through the thermo-optical coefficient $\frac{dn}{dT}$, leading to significant scattering, strong wavefront distortions, and ultimately poor Strehl ratios and low focused intensities. 
 
In this letter, we show experimentally how bringing the temperature of a weakly turbulent cooling liquid close to a point for which $\frac{dn}{dT}\simeq 0$, may result in the strong mitigation of turbulence-induced scattering thanks to index-leveling.
We discuss the potential to optimize the laser beam quality, or to push further the technological limits on high average-powers lasers.

\begin{figure}[htbp]
\centering
\includegraphics[width=\columnwidth]{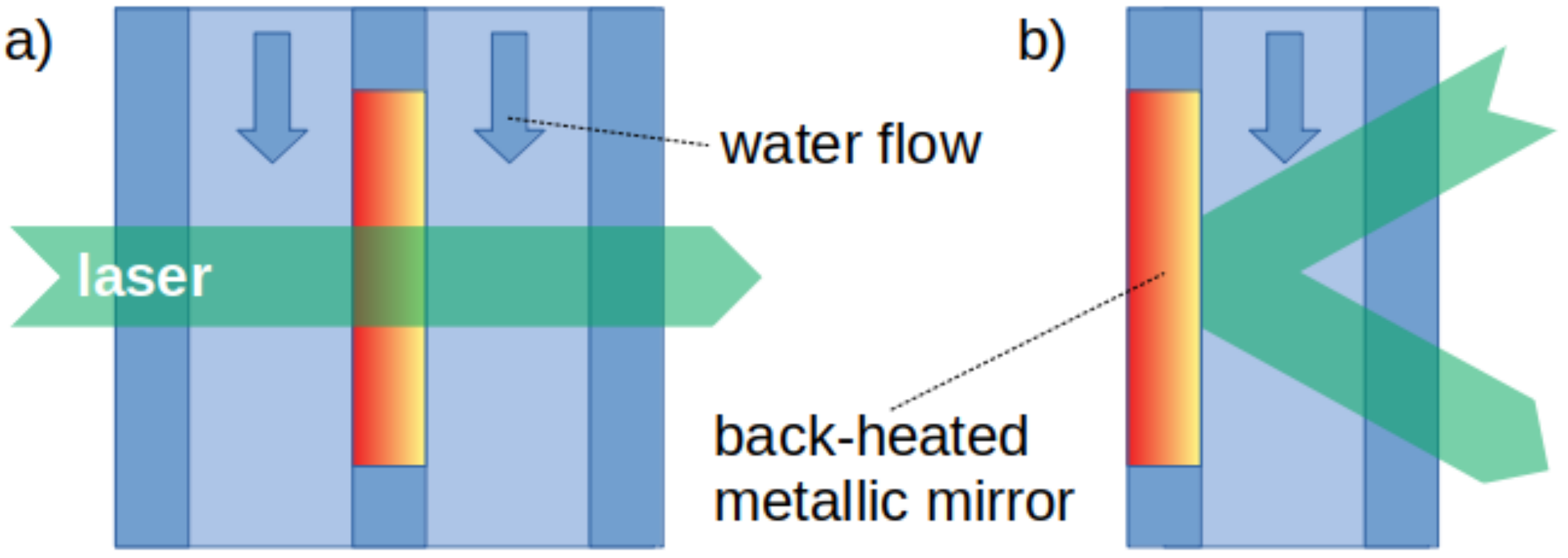}
\caption{\label{fig:sketch_slab_cooling}a) Liquid face cooling of a high power laser slab in a transmission geometry. b) Reflective ("active mirror") geometry used in the setup for analysis of wavefront perturbations.}
\end{figure}

Fig.~\ref{fig:sketch_slab_cooling} presents typical face-cooling mounts of a laser amplifier slab, in a straight configuration (a), and in a reflective configuration, such as used for "active mirrors" or thin-disks (b).
The coolant is sent into the head with a fixed overpressure given by an external hydraulic system. It is separated into two or more thin layers over the faces of the slab structure, which may be in a single slab or multi-slab configuration. The coolant canals therefore lie between an optical window and the slab, or between two slabs, with a separation $d$ and a transverse dimension $D$ usually exceeding the slab width. The coolant temperature is usually set above the condensation temperature of the room \cite{Koechner:06}.

The turbulence regime in each canal is specified by $\text{Re}=\frac{U_b \delta}{\nu},$
where $U_b$ is the mean coolant velocity, measured as the ratio of flow rate to canal area, $\nu$ its kinematic viscosity, and $\delta$ the hydraulic diameter for each canal, defined as four times the ratio between area and wet perimeters: $\delta =2\cdot d D / (d+D)$.

Our new approach mitigates wavefront deformations by operating the coolant close to a null thermo-optical point and at a high-Re transition regime. This concept stems from the progress in the understanding of free space optical beam communications, namely scintillation phenomena and loss of resolution power \cite{Kazaura:06}.
In complex transparent media, a laser beam goes through inhomogeneities of optical index, that, to first order, depend linearly on the local temperature and pressure:
\begin{equation}
\label{eq:n_taylor}
    n(T, P) =n_0+\frac{dn}{dT}dT + \frac{dn}{dP} dP.
\end{equation}

The pressure profile in the coolant flow is dominated by the constant hydrostatic pressure variation due to the external hydraulic system, along the flow direction, leading to a small  and constant optical prismaticity effect due to the $ \frac{dn}{dP}$ term. However, in each transverse plane, the pressure can be assumed constant in very good approximation  \cite{LivreJPHulin}, as all velocities within the flow are extremely small with respect to the sound velocity.
This leaves $T$ as the essential thermodynamic variable whose spatial fluctuations affect the optical index \cite{Hill:78}.
In a turbulent regime, the stochastic index inhomogeneities were shown to induce an average spectrum $\Phi_n$ of spatial frequencies $\kappa$ after propagation, that presents the following dependence for $\kappa$ larger than $\lambda ^{-1}$, where $\lambda$ is the laser wavelength:
\begin{equation}
\label{eq:turb_spatial_spectrum}
   \Phi_n\approx 0.23 \: C_n^2 \:  \eta ^{2/3}\: \kappa ^{-3}, 
\end{equation}
where $\eta$ is the Kolmogorov scale, and $C_n$ the so-called structure parameter of $n$ in the flow, dependent on $|dn/dT|$ and on the stochastic average of the temperature gradient $\left\langle\nabla T\right\rangle$ \cite{Hill:78,tatarski:16}.

\begin{figure}[htbp]
\centering\includegraphics[width=.8\columnwidth]{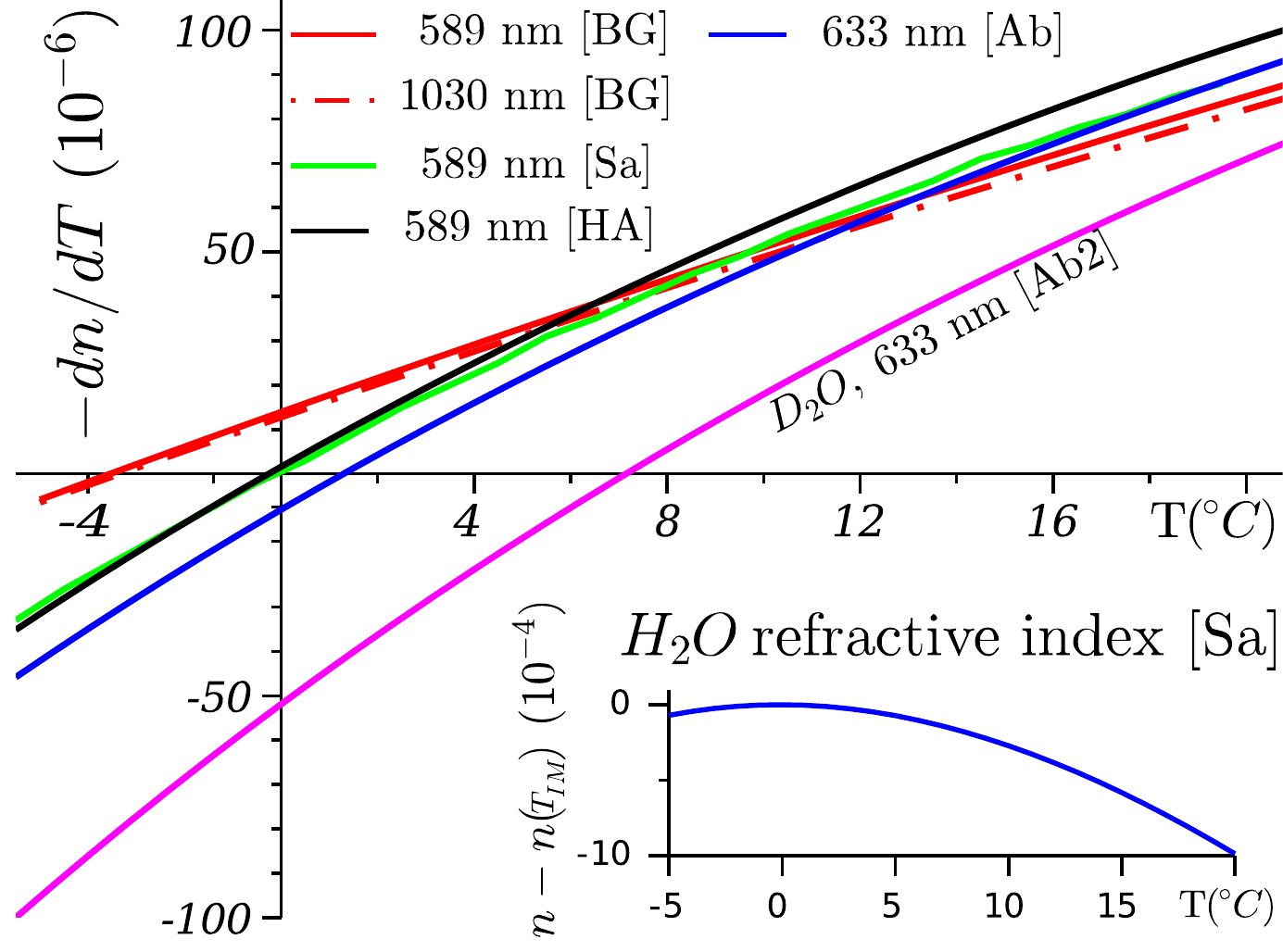}
\caption{\label{fig:biblio_opt_index_water}Optical index $n$ (inset) and thermo-optical coefficient $-\frac{dn}{dT}$ of H${}_2$O and D${}_2$O as measured by several authors. }
\end{figure}

$|dn/dT|$ thus appears as the key figure controlling the optical effects of turbulence. Fig.~\ref{fig:biblio_opt_index_water} illustrates the state of knowledge on the temperature dependence of the $dn/dT$ coefficient of pure water and of D${}_2$O.
The inset shows the behavior of the optical index of water, with a paraboloid shape around a temperature of index maximum $T_\text{IM}$ (data from \cite{Saubade:81}). The index presents a maximum around the melting point at ambient pressure. 
The main figure displays the fits for $-dn/dT$ from selected studies: Hawkes et al. \cite[HA]{Hawkes:48}, Saubade \cite[Sa]{Saubade:81},
 Abbate et al. \cite[Ab]{Abbate:78},
 and the  approximation formula for the H${}_2$O index of Bashkatov and Gelina \cite[BG]{Bashatov:03}.
The water's $T_\text{IM}$ lies in the range [-3, +2]~$^\circ$C depending on studies; the necessary use of overfusion water for these measurements is indeed delicate, explaining minor discrepancies in the literature.
 The dependence in $\lambda$ appears negligible between the visible and near-infrared ranges. The $T_\text{IM}$ of D${}_2$O is located around \SI[mode=text]{7}{\degreeCelsius} \cite[Ab2]{Abbate:80}.

\begin{figure}[htbp]
\centering\includegraphics[width=.75\columnwidth]{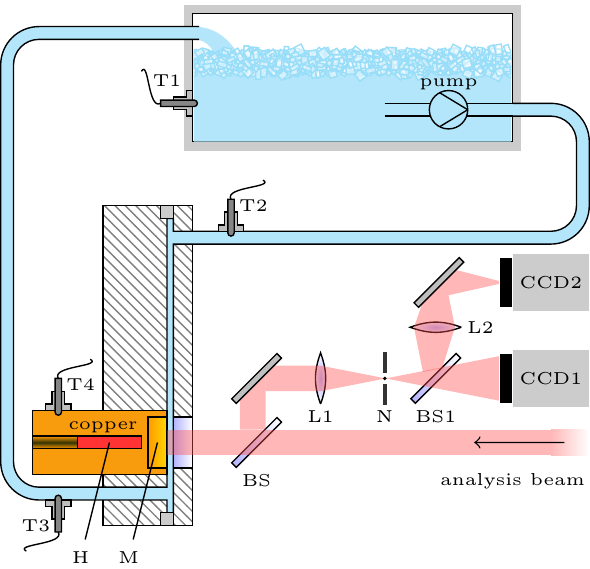}
\caption{\label{fig:exp_setup}Experimental setup used for the optical characterization of turbulence-induced wavefront perturbations.}
\end{figure}

We have chosen the reflective configuration of Fig.~\ref{fig:sketch_slab_cooling}(b) to test this proposal, and evaluate the extent to which temperature control around a null thermo-optical point allows one to mitigate wavefront corrugation. Fig.~\ref{fig:exp_setup} shows the experimental setup. 
Thanks to the reflective configuration, we could replace the hot laser crystal by a metallic mirror M heated from the back by an electrical resistance H, allowing us to tune the thermal power to extract, and hence the temperature difference between the optical surface and the incoming water coolant. 
The latter enters the mock-up amplifier head from the top, goes first into a 10-cm long turbulence decay section, then enters a 3-mm thick canal between the entrance window, and the mirror surface.
The flow follows a vertical top-down axis. The mirror itself is made of copper, with a high thermal conductivity ensuring an uniform temperature, and is layered with gold for good IR reflectivity. 
The electric power dissipated by H was regulated at \SI[mode=text]{150}{\W}.

\begin{figure}[htbp]
\centering\includegraphics[width=\columnwidth]{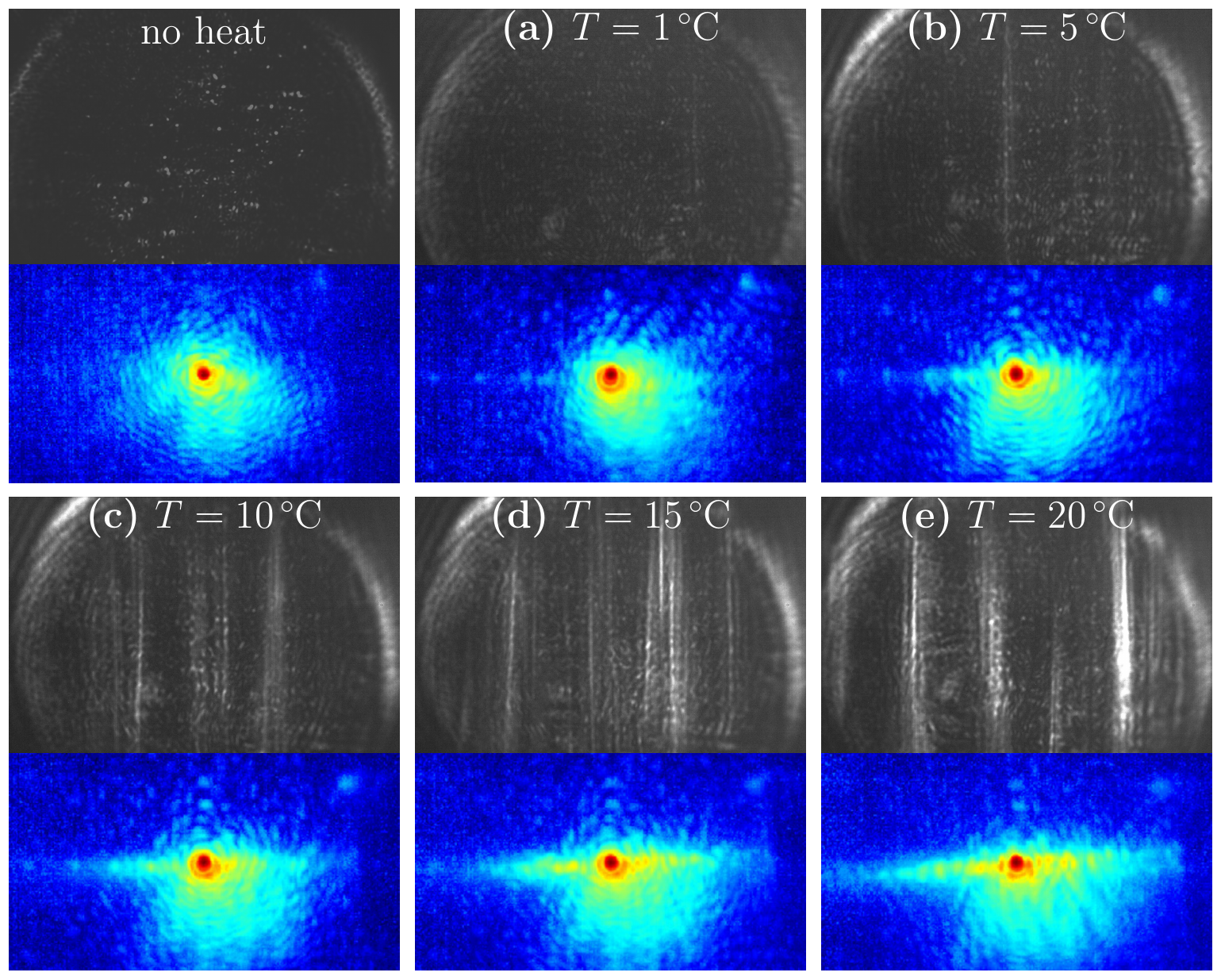}
\caption{\label{fig:results_near_far_field} For each set: strioscopic near-field (up), logarithm of far-field profiles (down) at $\text{Re}=4500$. Top left: without heating, a to e: 150~W heating, water temperature 1 to \SI[mode=text]{20}{\degreeCelsius}.}
\end{figure}

The water circulation is operated by an external hydraulic system comprising an hydraulic pump with adjustable overpressure, and a thermally isolated tank.
The water temperature is fixed by a thermostat between \SIlist[mode=text]{0;20}{\degreeCelsius}.
Indeed, as stated above, $T_\text{IM}$ lies close to the water's melting point. 
This thermodynamic coincidence makes the water thermostat technically challenging. To cool down the water very close to \SI[mode=text]{0}{\degreeCelsius}, we spilled ice chips directly in the water tank, letting it operate as a dual phase mixture, with the ice floating in the upper part, and the water derivation taken from the bottom of the tank.
The water temperature was measured in the tank by the sensor T1, and at the entrance and exit ports of the head by resp. T2 and T3, with a typical difference of \SI[mode=text]{0.3}{\degreeCelsius} in the conditions of Fig.~\ref{fig:results_near_far_field}. Sensor T4 was embedded in the copper close to M to monitor its temperature. All sensors (DS18B20) have a sensitivity of \SI[mode=text]{0.06}{\degreeCelsius}.
Temperatures below $\SI{10}{\degreeCelsius}$ usually induce condensation on the optical surfaces. We solved this issue by adding a transparent chamber (not shown) 
in front of the window, filled with dry air and dessicant.

A high spatial quality fiber laser was used to perform the optical qualification of the setup, displayed in the lower-right part of Fig.~\ref{fig:exp_setup}. We used a 10-mW source obtained by putting a 1-m Yb:glass active fiber in ring-cavity configuration, operating it as a spectrally-multimode laser centered at $\lambda = \SI{1030}{\nm}$.
The initially diverging beam is first collimated with a full width at half maximum of \SI[mode=text]{1}{\cm} and later impinges on the mock-up laser head after going through a beam splitter BS.
As a result, the laser beam goes back and forth through the entrance window and the coolant flow. On its way back, the beam is deflected by BS to an optical characterization setup, and split by the beam splitter BS1 into two optical imaging lines for the far- and near-fields.
 
In the latter, the laser profile at the mirror position is imaged onto the charge-coupled device camera CCD1 by means of a lens L1 with focal length $f_1 = \SI{50}{\cm}$. A needle N is inserted at the intermediate focus between L1 and CCD1. This gives a direct strioscopic image of index inhomogeneities  near the mirror M.

The second beamline uses L2 (focal length $f_2 = \SI{30}{\cm}$) to obtain a far-field profile at its focus, positioned at the object plane of the objective of CCD2, with the needle N removed.
The light scintillation induced by the turbulent fluid will translate into photons arriving in the focal plane away from the central laser spot, generating a wide spatial pedestal on the image.

Fig.~\ref{fig:results_near_far_field} shows a series of strioscopic near-field images (upper pictures), and of far-field high dynamic, logarithmic-scale images (lower pictures), taken at coolant temperatures of \SIlist[mode=text]{1;5;10;15;20}{\degreeCelsius} (a to e), along with a set of reference images taken without heat load at \SI[mode=text]{15}{\degreeCelsius}.
The hydrodynamic and thermal conditions were actively controlled to be identical for all temperatures, with $\text{Re}=4500$ and a temperature difference of \SI[mode=text]{17}{\degreeCelsius} between the mirror (sensor T4) and the coolant (sensor T2).
This required to adjust finely the flow rate to account for the temperature dependence of the kinematic viscosity. In such conditions, the turbulence regime was deemed identical for all data sets.

The near-field images show that the flow instabilities arise as a series of narrow threads, directed along the flow direction, and appearing randomly.
Somehow similar features can be noticed in \cite{Ruan:19}. Our pattern is much more regular in shape, possibly due to the geometry of the water canal, starting high above the mirror, bound to lead to a strictly parallel flow.
The laser scatters off these turbulence threads, leading to bright areas in strioscopic imaging. The diffraction pattern is thus bound to present a preferential direction orthogonal to the threads' axis, namely, along the horizontal direction.
This is indeed what is observed on the far-field log-scale images, with the appearance of comet-like horizontal features, superposed to the constant spatial pedestal of the beam, mainly due to surface irregularities of the metallic mirror.
Lengths and widths of the bright turbulence threads are similar throughout the series. Only their intensities fade away when the temperature approaches the thermo-optical null point, showing an almost complete optical bleaching of the turbulence structures.

\begin{figure}[htbp]
\centering\includegraphics[width=\columnwidth]{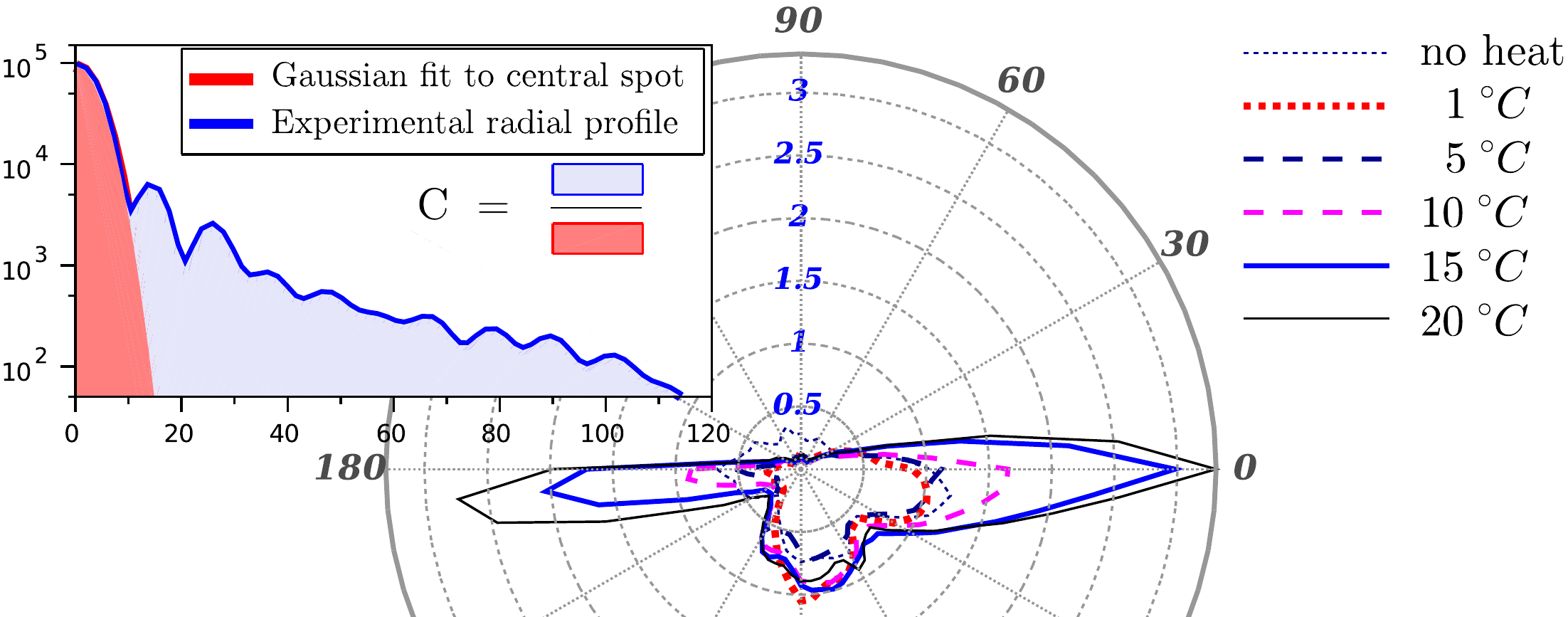}
\caption{\label{fig:polar_plot}Inset: graphical definition of the polar energy contrast ratio $C$. Polar plot: distributions of $C(\theta)$ for temperatures increasing from \SI[mode=text]{1}{\degreeCelsius} to \SI[mode=text]{20}{\degreeCelsius}.} 
\end{figure}

These qualitative considerations can be made quantitative by an analysis of the far-field data. In our numerical procedure \cite{Balcou:22}, we plot the energy profile of the far-field image in each polar direction, starting from the center of the focal spot, as shown in the inset of Fig.~\ref{fig:polar_plot}.
In logarithmic scale, the obtained curve displays a parabolic shape for small radii, characteristic of a Gaussian beam. The integral of the image intensity in this area provides the energy within the main focal spot. The curve at larger radii departs strongly from a parabola,  indicating the onset of the beam spatial pedestal that can be spatially integrated.
Defining the polar contrast $C$ as the ratio between pedestal energy (blue shading) and main beam energy (pink shading) in each angular sector $[\theta,\theta+d\theta]$ of the far-field profile, we can plot the polar distribution of the spatial contrast $C(\theta)$ as a function of water temperature, thus yielding a fully quantitative determination of the energy scattered off the beam, as shown in the polar plots of Fig.~\ref{fig:polar_plot}. For example, $C=1$ means that half of the beam energy is lost in the pedestal, whereas $C=0$ characterizes a quasi-Gaussian beam with only low-order aberrations.

The polar plots show  essentially constant profiles in the up and down directions, due to the intrinsic beam defects; plus sharp spikes, especially visible at \SI[mode=text]{15}{\degreeCelsius} and \SI[mode=text]{20}{\degreeCelsius}. In contrast, polar plots obtained at \SI[mode=text]{1}{\degreeCelsius} and \SI[mode=text]{5}{\degreeCelsius} are almost superposable.

\begin{figure}[htbp]
\centering\includegraphics[width=\columnwidth]{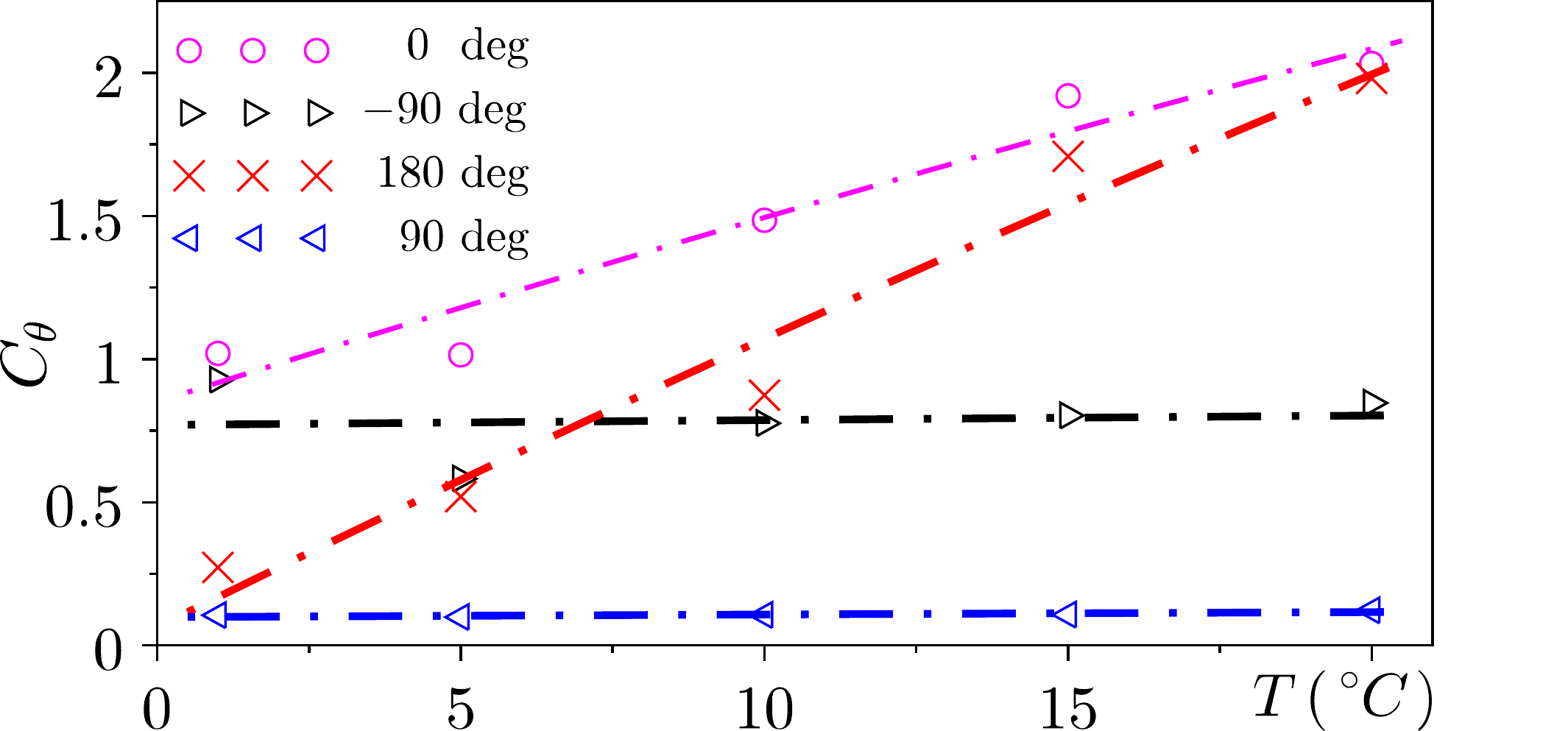}
\caption{\label{fig:polar_energy_ratios}Temperature dependence of $C(\theta)$ at cardinal directions.}
\end{figure}

Fig.~\ref{fig:polar_energy_ratios} shows the values of $C$ for the cardinal directions $\theta  = \SI{0}{\degree}$ (circles), $\theta  = \pm \SI{90}{\degree}$ ((left and right pointing arrows), and $\theta \simeq \SI{180}{\degree}$ (crosses, along the light tail).
Up and down contrast ratios are simply constant; lateral contrast ratios exhibit quasi-linear increases. The dashed curves are numerical regressions of the form $C=\alpha + \beta \frac{dn}{dT} (T)$, with $\frac{dn}{dT}(T)$ following the formula derived in \cite{Bashatov:03} (cf. Fig.~\ref{fig:biblio_opt_index_water}). The improvement of $C(\theta)$ appears spectacular at \SI[mode=text]{0}{\degree} and \SI[mode=text]{180}{\degree}. $C_{180}$ is eg. decreased by a factor 7 between \SI[mode=text]{20}{\degreeCelsius} and \SI[mode=text]{1}{\degreeCelsius}, a value limited by the pre-existing static spatial pedestal visible on CCD2 without heating (cf. Fig.~\ref{fig:results_near_far_field}, top left set).

In usual geometries, the heat transfer coefficient scales as $\text{Re}^{0.5}$ in the laminar regime \cite{churchill:73} and $\text{Re}^{0.8}$ in the turbulent regime \cite{Koechner:06}.
While Re in the literature rarely exceeds 1100 for immersed amplifiers \cite{Ruan:19} to maintain an acceptable beam quality, the present study demonstrates the optical viability of the index-leveling method at least up to $\text{Re} = 4500$, that should hence lead to a several-fold mitigation of the temperature rise in face-cooled optics.
Larger Re values should be investigated in the future.

Our simple setup already dissipates \SI[mode=text]{150}{\W} on a 1-inch single surface.
To put this cooling capacity in perspective, we may take the example of a Yb:YAG triple-slab amplifier in a straight geometry \cite{lhermite:21}, pumped at its zero-phonon line ($\lambda = \SI{970}{\nm}$).
A power of $6\times\SI{150}{\W} = \SI{900}{\W}$ represents the local, non-radiative losses associated to the quantum defect of the 1030-nm emission line of Yb$^{3+}$ when pumped at a power level of \SI[mode=text]{15}{\kW}.
With a typical efficiency of 15\% in pulsed mode, such an amplifier may output a power level of \SI[mode=text]{2300}{\W}.
This simple evaluation shows the potential of the method to boost the average power of intense lasers beyond the kilowatt, provided the issues of condensation and chiller design can be dealt with at that scale. However, as hinted by Fig.~\ref{fig:biblio_opt_index_water}, using D$_2$O at \SI[mode=text]{7}{\degreeCelsius} would be very helpful on these issues, even if costly, with the additional benefit of reduced laser and pump absorption, as its extinction length is about ten times higher than that of H$_2$O in the near-IR range.

The heat exchange capacity demonstrated here appears already superior to %that of
room-temperature He gas cooling \cite{Bayramian:07}. Further studies are required to evaluate the advantages and drawbacks of index-leveling with respect to other technologies such as thin disks and cryogenic gas cooling \cite{divoky:21}. Its potential should in particular be investigated for large laser slabs used in frontier kilojoule high-repetition lasers, such as the ELI Beamlines \cite{Gaul:18}.

In conclusion, the proposed index-leveling method using a null thermo-optic point allows to operate face-cooled laser amplifiers in a new high-Re, weakly turbulent regime, thus enabling stronger exchange coefficients and more efficient cooling.
In such conditions, the beam scattering over the turbulence threads is effectively suppressed, as the optical index becomes independent of local temperatures to first order.
As such, the method may help to simplify the design of high average-power amplifier heads, and ultimately open up a host of new laser applications from satellite deorbitation to medical X-ray generation.

%\section*{Acknowledgments}
This work is part of the LEAP project, supported by Région Aquitaine (contract 16005645), European Structural Funds (contract 2779610), Université de Bordeaux and CNRS. Discussions with D. Descamps and J.-P. Hulin are gratefully acknowledged.

%%%%%%%%%%% References %%%%%%%%%%%%%%%%%%%
\bibliography{index-leveling_cooling}

\begin{thebibliography}{24}
\expandafter\ifx\csname natexlab\endcsname\relax\def\natexlab#1{#1}\fi
\expandafter\ifx\csname bibnamefont\endcsname\relax
  \def\bibnamefont#1{#1}\fi
\expandafter\ifx\csname bibfnamefont\endcsname\relax
  \def\bibfnamefont#1{#1}\fi
\expandafter\ifx\csname citenamefont\endcsname\relax
  \def\citenamefont#1{#1}\fi
\expandafter\ifx\csname url\endcsname\relax
  \def\url#1{\texttt{#1}}\fi
\expandafter\ifx\csname urlprefix\endcsname\relax\def\urlprefix{URL }\fi
\providecommand{\bibinfo}[2]{#2}
\providecommand{\eprint}[2][]{\url{#2}}

\bibitem[{\citenamefont{Richardson et~al.}(2010)\citenamefont{Richardson,
  Nilsson, and Clarkson}}]{Richardson:10}
\bibinfo{author}{\bibfnamefont{D.~J.} \bibnamefont{Richardson}},
  \bibinfo{author}{\bibfnamefont{J.}~\bibnamefont{Nilsson}}, \bibnamefont{and}
  \bibinfo{author}{\bibfnamefont{W.~A.} \bibnamefont{Clarkson}},
  \bibinfo{journal}{J. Opt. Soc. Am. B} \textbf{\bibinfo{volume}{27}}
  (\bibinfo{year}{2010}).

\bibitem[{\citenamefont{Giesen and Speiser}(2007)}]{Giesen:07}
\bibinfo{author}{\bibfnamefont{A.}~\bibnamefont{Giesen}} \bibnamefont{and}
  \bibinfo{author}{\bibfnamefont{J.}~\bibnamefont{Speiser}},
  \bibinfo{journal}{IEEE Journal of Selected Topics in Quantum Electronics}
  \textbf{\bibinfo{volume}{13}}, \bibinfo{pages}{598} (\bibinfo{year}{2007}).

\bibitem[{\citenamefont{Albach et~al.}(2019)\citenamefont{Albach, Loeser,
  Siebold, and Schramm}}]{Albach:19}
\bibinfo{author}{\bibfnamefont{D.}~\bibnamefont{Albach}},
  \bibinfo{author}{\bibfnamefont{M.}~\bibnamefont{Loeser}},
  \bibinfo{author}{\bibfnamefont{M.}~\bibnamefont{Siebold}}, \bibnamefont{and}
  \bibinfo{author}{\bibfnamefont{U.}~\bibnamefont{Schramm}},
  \bibinfo{journal}{High Power Laser Science and Engineering}
  \textbf{\bibinfo{volume}{7}} (\bibinfo{year}{2019}).

\bibitem[{\citenamefont{Lin et~al.}(2017)\citenamefont{Lin, Zhu, Zhao, Qiao,
  Wang, Wang, and Zhu}}]{Lin:17}
\bibinfo{author}{\bibfnamefont{Z.}~\bibnamefont{Lin}},
  \bibinfo{author}{\bibfnamefont{G.}~\bibnamefont{Zhu}},
  \bibinfo{author}{\bibfnamefont{W.}~\bibnamefont{Zhao}},
  \bibinfo{author}{\bibfnamefont{Y.}~\bibnamefont{Qiao}},
  \bibinfo{author}{\bibfnamefont{M.}~\bibnamefont{Wang}},
  \bibinfo{author}{\bibfnamefont{H.}~\bibnamefont{Wang}}, \bibnamefont{and}
  \bibinfo{author}{\bibfnamefont{X.}~\bibnamefont{Zhu}}, \bibinfo{journal}{J.
  Opt. Soc. Am. B} \textbf{\bibinfo{volume}{34}}, \bibinfo{pages}{1669}
  (\bibinfo{year}{2017}).

\bibitem[{\citenamefont{Okada et~al.}(2006)\citenamefont{Okada, Yoshida,
  Fujita, and Nakatsuka}}]{Okada:06}
\bibinfo{author}{\bibfnamefont{H.}~\bibnamefont{Okada}},
  \bibinfo{author}{\bibfnamefont{H.}~\bibnamefont{Yoshida}},
  \bibinfo{author}{\bibfnamefont{H.}~\bibnamefont{Fujita}}, \bibnamefont{and}
  \bibinfo{author}{\bibfnamefont{M.}~\bibnamefont{Nakatsuka}},
  \bibinfo{journal}{Opt. Comm.} \textbf{\bibinfo{volume}{266}}
  (\bibinfo{year}{2006}).

\bibitem[{\citenamefont{Lhermite et~al.}(2021)\citenamefont{Lhermite, Féral,
  Marion, Rohm, Balcou, Descamps, Petit, Nadeau, and Mével}}]{lhermite:21}
\bibinfo{author}{\bibfnamefont{J.}~\bibnamefont{Lhermite}},
  \bibinfo{author}{\bibfnamefont{{\relax Ch}.}~\bibnamefont{Féral}},
  \bibinfo{author}{\bibfnamefont{D.}~\bibnamefont{Marion}},
  \bibinfo{author}{\bibfnamefont{A.}~\bibnamefont{Rohm}},
  \bibinfo{author}{\bibfnamefont{{\relax Ph}.}~\bibnamefont{Balcou}},
  \bibinfo{author}{\bibfnamefont{D.}~\bibnamefont{Descamps}},
  \bibinfo{author}{\bibfnamefont{S.}~\bibnamefont{Petit}},
  \bibinfo{author}{\bibfnamefont{{\relax M.-Ch.}.}~\bibnamefont{Nadeau}},
  \bibnamefont{and} \bibinfo{author}{\bibfnamefont{E.}~\bibnamefont{Mével}},
  in \emph{\bibinfo{booktitle}{International Conference on Advanced Laser
  Technologies (ALT)}} (\bibinfo{organization}{Prokhorov General Physics
  Institute of Russian Academics of Sciences}, \bibinfo{year}{2021}),
  \bibinfo{number}{21}, p. \bibinfo{pages}{133}.

\bibitem[{\citenamefont{Siebold et~al.}(2014)\citenamefont{Siebold, Loeser,
  Harzendorf, Nehring, Tsybin, Roeser, Albach, and Schramm}}]{Siebold:14}
\bibinfo{author}{\bibfnamefont{M.}~\bibnamefont{Siebold}},
  \bibinfo{author}{\bibfnamefont{M.}~\bibnamefont{Loeser}},
  \bibinfo{author}{\bibfnamefont{G.}~\bibnamefont{Harzendorf}},
  \bibinfo{author}{\bibfnamefont{H.}~\bibnamefont{Nehring}},
  \bibinfo{author}{\bibfnamefont{I.}~\bibnamefont{Tsybin}},
  \bibinfo{author}{\bibfnamefont{F.}~\bibnamefont{Roeser}},
  \bibinfo{author}{\bibfnamefont{D.}~\bibnamefont{Albach}}, \bibnamefont{and}
  \bibinfo{author}{\bibfnamefont{U.}~\bibnamefont{Schramm}},
  \bibinfo{journal}{Optics Letters} \textbf{\bibinfo{volume}{39}},
  \bibinfo{pages}{3611} (\bibinfo{year}{2014}).

\bibitem[{\citenamefont{Hill}(1978)}]{Hill:78}
\bibinfo{author}{\bibfnamefont{R.}~\bibnamefont{Hill}}, \bibinfo{journal}{J.
  Opt. Soc. Am.} \textbf{\bibinfo{volume}{68}}, \bibinfo{pages}{1067}
  (\bibinfo{year}{1978}).

\bibitem[{\citenamefont{Hill and Clifford}(1981)}]{Hill:81}
\bibinfo{author}{\bibfnamefont{R.}~\bibnamefont{Hill}} \bibnamefont{and}
  \bibinfo{author}{\bibfnamefont{S.}~\bibnamefont{Clifford}},
  \bibinfo{journal}{J. Opt. Soc. Am.} \textbf{\bibinfo{volume}{71}},
  \bibinfo{pages}{675} (\bibinfo{year}{1981}).

\bibitem[{\citenamefont{Koechner}(2006)}]{Koechner:06}
\bibinfo{author}{\bibfnamefont{W.}~\bibnamefont{Koechner}},
  \emph{\bibinfo{title}{Solid State Laser Engineering}}
  (\bibinfo{publisher}{Springer}, \bibinfo{year}{2006}), \bibinfo{edition}{6th}
  ed.

\bibitem[{\citenamefont{Kazaura et~al.}(2006)\citenamefont{Kazaura, Omae,
  Suzuki, Matsumoto, Mutafungwa, Korhonen, Murakami, Takahashi, Matsumoto,
  Wakamori et~al.}}]{Kazaura:06}
\bibinfo{author}{\bibfnamefont{K.}~\bibnamefont{Kazaura}},
  \bibinfo{author}{\bibfnamefont{K.}~\bibnamefont{Omae}},
  \bibinfo{author}{\bibfnamefont{T.}~\bibnamefont{Suzuki}},
  \bibinfo{author}{\bibfnamefont{M.}~\bibnamefont{Matsumoto}},
  \bibinfo{author}{\bibfnamefont{E.}~\bibnamefont{Mutafungwa}},
  \bibinfo{author}{\bibfnamefont{T.~O.} \bibnamefont{Korhonen}},
  \bibinfo{author}{\bibfnamefont{T.}~\bibnamefont{Murakami}},
  \bibinfo{author}{\bibfnamefont{K.}~\bibnamefont{Takahashi}},
  \bibinfo{author}{\bibfnamefont{H.}~\bibnamefont{Matsumoto}},
  \bibinfo{author}{\bibfnamefont{K.}~\bibnamefont{Wakamori}},
  \bibnamefont{et~al.}, \bibinfo{journal}{Opt. Express}
  \textbf{\bibinfo{volume}{14}}, \bibinfo{pages}{4958} (\bibinfo{year}{2006}).

\bibitem[{\citenamefont{Guyon et~al.}(2015)\citenamefont{Guyon, Hulin, Petit,
  and Mitescu}}]{LivreJPHulin}
\bibinfo{author}{\bibfnamefont{E.}~\bibnamefont{Guyon}},
  \bibinfo{author}{\bibfnamefont{J.-P.} \bibnamefont{Hulin}},
  \bibinfo{author}{\bibfnamefont{L.}~\bibnamefont{Petit}}, \bibnamefont{and}
  \bibinfo{author}{\bibfnamefont{C.~D.} \bibnamefont{Mitescu}},
  \emph{\bibinfo{title}{Physical hydrodynamics}} (\bibinfo{publisher}{Oxford
  University Press}, \bibinfo{year}{2015}).

\bibitem[{\citenamefont{Tatarski}(2016)}]{tatarski:16}
\bibinfo{author}{\bibfnamefont{V.~I.} \bibnamefont{Tatarski}},
  \emph{\bibinfo{title}{Wave propagation in a turbulent medium}}
  (\bibinfo{publisher}{Courier Dover Publications}, \bibinfo{year}{2016}),
  \bibinfo{note}{~pp.~97--102}.

\bibitem[{\citenamefont{Saubade}(1981)}]{Saubade:81}
\bibinfo{author}{\bibfnamefont{C.}~\bibnamefont{Saubade}}, \bibinfo{journal}{J.
  Phys. France} \textbf{\bibinfo{volume}{42}}, \bibinfo{pages}{359}
  (\bibinfo{year}{1981}).

\bibitem[{\citenamefont{Hawkes and Astheimer}(1948)}]{Hawkes:48}
\bibinfo{author}{\bibfnamefont{J.}~\bibnamefont{Hawkes}} \bibnamefont{and}
  \bibinfo{author}{\bibfnamefont{R.}~\bibnamefont{Astheimer}},
  \bibinfo{journal}{J. Opt. Soc. Am.} \textbf{\bibinfo{volume}{38}},
  \bibinfo{pages}{804} (\bibinfo{year}{1948}).

\bibitem[{\citenamefont{Abbate et~al.}(1978)\citenamefont{Abbate, Bernini,
  Ragozzino, and Somma}}]{Abbate:78}
\bibinfo{author}{\bibfnamefont{G.}~\bibnamefont{Abbate}},
  \bibinfo{author}{\bibfnamefont{U.}~\bibnamefont{Bernini}},
  \bibinfo{author}{\bibfnamefont{E.}~\bibnamefont{Ragozzino}},
  \bibnamefont{and} \bibinfo{author}{\bibfnamefont{F.}~\bibnamefont{Somma}},
  \bibinfo{journal}{Journal of Physics D: Applied Physics}
  \textbf{\bibinfo{volume}{11}}, \bibinfo{pages}{1167} (\bibinfo{year}{1978}).

\bibitem[{\citenamefont{Bashkatov and Genina}(2003)}]{Bashatov:03}
\bibinfo{author}{\bibfnamefont{A.~N.} \bibnamefont{Bashkatov}}
  \bibnamefont{and} \bibinfo{author}{\bibfnamefont{E.~A.}
  \bibnamefont{Genina}}, in \emph{\bibinfo{booktitle}{Saratov Fall Meeting
  2002: Optical Technologies in Biophysics and Medicine IV}}
  (\bibinfo{organization}{International Society for Optics and Photonics},
  \bibinfo{year}{2003}), vol. \bibinfo{volume}{5068}, pp.
  \bibinfo{pages}{393--395}.

\bibitem[{\citenamefont{Abbate et~al.}(1980)\citenamefont{Abbate, Bernini,
  Ragozzino, and Somma}}]{Abbate:80}
\bibinfo{author}{\bibfnamefont{G.}~\bibnamefont{Abbate}},
  \bibinfo{author}{\bibfnamefont{U.}~\bibnamefont{Bernini}},
  \bibinfo{author}{\bibfnamefont{E.}~\bibnamefont{Ragozzino}},
  \bibnamefont{and} \bibinfo{author}{\bibfnamefont{F.}~\bibnamefont{Somma}},
  \bibinfo{journal}{Z. Naturforsch. A} \textbf{\bibinfo{volume}{35a}},
  \bibinfo{pages}{1171} (\bibinfo{year}{1980}).

\bibitem[{\citenamefont{Ruan et~al.}(2019)\citenamefont{Ruan, Su, Tu, Shang,
  Wu, Yi, Cao, Ma, Wang, Shen et~al.}}]{Ruan:19}
\bibinfo{author}{\bibfnamefont{X.}~\bibnamefont{Ruan}},
  \bibinfo{author}{\bibfnamefont{H.}~\bibnamefont{Su}},
  \bibinfo{author}{\bibfnamefont{B.}~\bibnamefont{Tu}},
  \bibinfo{author}{\bibfnamefont{J.}~\bibnamefont{Shang}},
  \bibinfo{author}{\bibfnamefont{J.}~\bibnamefont{Wu}},
  \bibinfo{author}{\bibfnamefont{J.}~\bibnamefont{Yi}},
  \bibinfo{author}{\bibfnamefont{H.}~\bibnamefont{Cao}},
  \bibinfo{author}{\bibfnamefont{Y.}~\bibnamefont{Ma}},
  \bibinfo{author}{\bibfnamefont{G.}~\bibnamefont{Wang}},
  \bibinfo{author}{\bibfnamefont{D.}~\bibnamefont{Shen}}, \bibnamefont{et~al.},
  \bibinfo{journal}{Optics Communications} \textbf{\bibinfo{volume}{436}},
  \bibinfo{pages}{26} (\bibinfo{year}{2019}).

\bibitem[{\citenamefont{Balcou et~al.}(2022)\citenamefont{Balcou, Marion, Rohm,
  Féral, and Lhermite}}]{Balcou:22}
\bibinfo{author}{\bibfnamefont{{\relax Ph}.}~\bibnamefont{Balcou}},
  \bibinfo{author}{\bibfnamefont{D.}~\bibnamefont{Marion}},
  \bibinfo{author}{\bibfnamefont{A.}~\bibnamefont{Rohm}},
  \bibinfo{author}{\bibfnamefont{{\relax Ch}.}~\bibnamefont{Féral}},
  \bibnamefont{and} \bibinfo{author}{\bibfnamefont{J.}~\bibnamefont{Lhermite}},
  \bibinfo{journal}{to be published}  (\bibinfo{year}{2022}).

\bibitem[{\citenamefont{Churchill and Ozoe}(1973)}]{churchill:73}
\bibinfo{author}{\bibfnamefont{S.~W.} \bibnamefont{Churchill}}
  \bibnamefont{and} \bibinfo{author}{\bibfnamefont{H.}~\bibnamefont{Ozoe}},
  \bibinfo{journal}{J. Heat Transfer} \textbf{\bibinfo{volume}{95}},
  \bibinfo{pages}{573} (\bibinfo{year}{1973}).

\bibitem[{\citenamefont{Bayramian et~al.}(2007)\citenamefont{Bayramian,
  Armstrong, and Ault}}]{Bayramian:07}
\bibinfo{author}{\bibfnamefont{A.}~\bibnamefont{Bayramian}},
  \bibinfo{author}{\bibfnamefont{P.}~\bibnamefont{Armstrong}},
  \bibnamefont{and} \bibinfo{author}{\bibfnamefont{E.}~\bibnamefont{Ault}},
  \bibinfo{journal}{Fusion Sc. and \& Tech.} \textbf{\bibinfo{volume}{52}},
  \bibinfo{pages}{383} (\bibinfo{year}{2007}).

\bibitem[{\citenamefont{Divoký et~al.}(2021)\citenamefont{Divoký, Pilař,
  Hanuš, Navrátil, Denk, Severová, Mason, Butcher, Banerjee, Vido
  et~al.}}]{divoky:21}
\bibinfo{author}{\bibfnamefont{M.}~\bibnamefont{Divoký}},
  \bibinfo{author}{\bibfnamefont{J.}~\bibnamefont{Pilař}},
  \bibinfo{author}{\bibfnamefont{M.}~\bibnamefont{Hanuš}},
  \bibinfo{author}{\bibfnamefont{P.}~\bibnamefont{Navrátil}},
  \bibinfo{author}{\bibfnamefont{O.}~\bibnamefont{Denk}},
  \bibinfo{author}{\bibfnamefont{P.}~\bibnamefont{Severová}},
  \bibinfo{author}{\bibfnamefont{P.}~\bibnamefont{Mason}},
  \bibinfo{author}{\bibfnamefont{T.}~\bibnamefont{Butcher}},
  \bibinfo{author}{\bibfnamefont{S.}~\bibnamefont{Banerjee}},
  \bibinfo{author}{\bibfnamefont{M.~D.} \bibnamefont{Vido}},
  \bibnamefont{et~al.}, \bibinfo{journal}{Opt. Lett.}
  \textbf{\bibinfo{volume}{46}}, \bibinfo{pages}{5771} (\bibinfo{year}{2021}).

\bibitem[{\citenamefont{Gaul et~al.}(2018)\citenamefont{Gaul, Chériaux,
  Antipenkov, Batysta, Borger, Friedman, Greene, Hammond, Heisler, Hidinger
  et~al.}}]{Gaul:18}
\bibinfo{author}{\bibfnamefont{E.}~\bibnamefont{Gaul}},
  \bibinfo{author}{\bibfnamefont{G.}~\bibnamefont{Chériaux}},
  \bibinfo{author}{\bibfnamefont{R.}~\bibnamefont{Antipenkov}},
  \bibinfo{author}{\bibfnamefont{F.}~\bibnamefont{Batysta}},
  \bibinfo{author}{\bibfnamefont{T.}~\bibnamefont{Borger}},
  \bibinfo{author}{\bibfnamefont{G.}~\bibnamefont{Friedman}},
  \bibinfo{author}{\bibfnamefont{J.}~\bibnamefont{Greene}},
  \bibinfo{author}{\bibfnamefont{D.}~\bibnamefont{Hammond}},
  \bibinfo{author}{\bibfnamefont{J.}~\bibnamefont{Heisler}},
  \bibinfo{author}{\bibfnamefont{D.}~\bibnamefont{Hidinger}},
  \bibnamefont{et~al.}, in \emph{\bibinfo{booktitle}{Conference on Lasers and
  Electro-Optics}} (\bibinfo{organization}{Optica Publishing Group},
  \bibinfo{year}{2018}), p. \bibinfo{pages}{STu3M.2}.

\end{thebibliography}

\end{document}